# Reduced integration and improved segregation of functional brain networks in Alzheimer's disease

Kabbara A.[1, 2, 3], Eid H.[4], EL Falou W.[2, 3], Khalil M.[2, 3], Wendling F.[1 ϕ], Hassan M.[1 * ϕ]

[1] Univ Rennes, LTSI, F-35000 Rennes, France

[2] Azm Center for Research in Biotechnology and its Application, EDST, Lebanese University, Lebanon

[3] CRSI research center, Faculty of Engineering, Lebanese University, Lebanon

[4] Mazloum Hospital, Tripoli, Lebanon

[ϕ] These authors contributed equally to this work

[*] Corresponding author: mahmoud.hassan@univ-rennes1.fr

# Abstract

*Objective:* Emerging evidence shows that cognitive deficits in Alzheimer's disease (AD) are associated with disruptions in brain functional connectivity. Thus, the identification of alterations in AD functional networks has become a topic of increasing interest. However, to what extent AD induces disruption of the balance of local and global information processing in the human brain remains elusive. The main objective of this study is to explore the dynamic topological changes of AD networks in terms of brain network segregation and integration.

*Approach:* We used electroencephalography (EEG) data recorded from 20 participants (10 AD patients and 10 healthy controls) during resting state. Functional brain networks were reconstructed using EEG source connectivity computed in different frequency bands. Graph theoretical analyses were performed assess differences between both groups.

*Main results:* Results revealed that AD networks, compared to networks of age-matched healthy controls, are characterized by lower global information processing (integration) and higher local information processing (segregation). Results showed also significant correlation between the alterations in the AD patients' functional brain networks and their cognitive scores.

*Significance:* These findings may contribute to the development of EEG network-based test that could strengthen results obtained from currently-used neurophysiological tests in neurodegenerative diseases.

# Introduction

Worldwide, about 35 million people are estimated to have dementia (World Health Organization, 2012). Alzheimer's disease (AD), the most common cause of dementia, is a neurological disorder essentially characterized by progressive impairment of memory and other cognitive functions. Emerging evidence show that the progressive evolution in AD is related to pathological changes in large-scale networks (Supekar *et al.*, 2008; Zhou *et al.*, 2010; Pievani *et al.*, 2011). Therefore, from a clinical perspective, the demand is high for non-invasive and easy-to-use methods to identify pathological alterations in brain networks. More precisely, novel 'neuromarkers' able to identify and characterize networks associated with cognitive deficits in AD patients, in particular at early stage, are needed.

In this context, electroencephalography (EEG) has some major assets since it is a non-invasive, easy to use and clinically available technique. A potential framework for advanced EEG analysis is the emerging technique called "MEG/EEG source connectivity" (de Pasquale *et al.*, 2010; Hipp *et al.*, 2012; Mehrkanoon *et al.*, 2014; Hassan *et al.*, 2015; Kabbara *et al.*, 2017). As shown by several recent studies (Hassan *et al.*, 2016, 2017; Engels *et al.*, 2017), this technique could indeed respond to clinical demand, provided that appropriate information processing is performed. Previous results, using the EEG source connectivity methods, showed alterations in the functional connectivity at the theta and alpha2 bands in AD patients compared to controls (Canuet *et al.*, 2012). Relationships between the dysfunctional connections in AD patients and the cognitive decline progression were also observed (Hata *et al.*, 2016). Moreover, Vecchio et al. showed, in a large group of AD patients, changes in topological brain network characteristics mainly in the clustering coefficient and the path length measures (Vecchio *et al.*, 2014).

However, to what extent the AD modifies the brain network segregation (local information processing) and integration (global information processing) remains unclear. This is the main objective of the paper. More precisely, we address two questions: i) do the dynamic brain network segregation and integration changes in AD compared to controls? And ii) is there a correlation between the network disruptions and the cognitive score of the AD patients? To tackle this issue, we combined the use of the EEG source connectivity with the graph theory based analysis. Resting state EEG data were recorded from 20 participants (10 AD patients and 10 age-matched controls). The functional networks were reconstructed at the cortical level from scalp EEG electrodes. The identified networks were then analyzed by graph measures that allow the characterization of these networks at different scales from high-level topology to low-level topology.

# Materials and methods

The full pipeline of this study is illustrated in Figure 1.

**Participants**

Ten healthy controls (6 males and 4 females, age 64-78 y) and ten patients diagnosed with AD (5 females and 5 males, age 66–81 y) participated in this study. All subjects provided informed consent in accordance with the local institutional review boards guidelines (CE-EDST-3-2017). Patients were recruited from the memory clinic of Dar al-Ajaza Hospital and from Mazloum Hospital, Tripoli, Lebanon. Age-matched healthy controls were recruited from Dar Al-Ajaza Hospital and the local community. For each subject medical history, a cognitive screening test and EEG recording were performed. The mini-mental state examination (MMSE) was used as an indicator of the global cognitive performance (Folstein, Folstein and McHugh, 1975). This test has been widely used to characterize the overall cognitive level of AD patients and to estimate the severity and progression of cognitive impairment (Ismail, Rajji and Shulman, 2010). Based on (Mungas, 1991), any score greater than or equal to 24 points out of 30 (MMSE $\geq$ 0.8) indicates normal cognitive functions. Below this score indicate cognitive impairment.

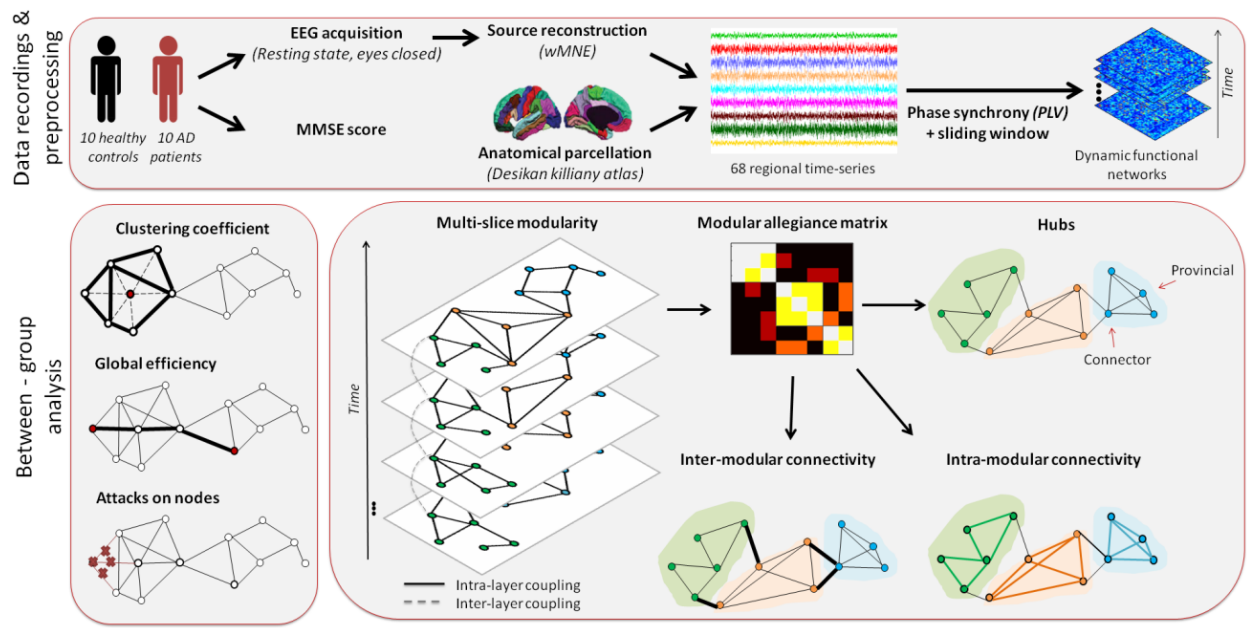

**Figure 1.** *Design of the study*. Data were recorded from 10 healthy controls and 10 AD patients during resting state condition (eyes closed). The cognitive performance was evaluated using MMSE score. The cortical sources were reconstructed using weighted minimum norm estimate (wMNE) inverse solution. Desikan Killiany atlas was used to anatomically parcellate the brain into 68 ROIs. The dynamic functional networks were then computed using phase synchrony method combined with a sliding window approach. In order to analyze the difference between healthy and AD networks, graph measures were extracted: clustering coefficient, global efficiency and vulnerability of each node (influence of each node's attack on the network global efficiency). Modularity-based parameters (mainly integration and segregation of networks) were also used. Moreover, the network hubs of each group were identified and compared.

## Data acquisition and preprocessing

EEG signals were recorded using a 32-channel EEG system (Twente Medical Systems International -TMSi-, Porti system) placed on the head according to the 10-20 system (Klem *et al.*, 1958). Signals were sampled at 500 Hz and band-pass filtered between 0.1-45 Hz. All subjects underwent 10 min of resting-state in which they were asked to relax and keep their eyes closed without falling asleep.

EEG signals are often contaminated by several sources of noise and artifacts. In order to clean raw signals, the pre-processing followed the same steps as described in several previous studies dealing with EEG resting state data (Onton *et al.*, 2006; Korjus *et al.*, 2015; Li *et al.*, 2015; Hassan *et al.*, 2017; Kabbara *et al.*, 2017). Briefly, the bad channels (i.e displaying signals that are either completely flat or are contaminated by movement artifacts) were first identified by visual inspection, complemented by the power spectral density, when needed. Then, these bad channels were recovered using an spherical interpolation procedure implemented in EEGLAB (Delorme and Makeig, 2004). In addition, epochs with voltage fluctuation >+80 µV and <-80 µV were removed. Consequently, for each participant, four artifact-free epochs of 40s lengths were selected. This epoch length was largely used previously and considered as a good compromise between the needed temporal resolution and the reproducibility of the results (Kabbara *et al.*, 2017). As the recorded EEG data used here has a very high temporal resolution (~1ms), the number of available samples is largely sufficient to compute statistically-consistent functional networks. By using a sliding window approach while calculating the functional connectivity, a high number of networks were obtained for each 40s-epoch and for different frequency bands.

The EEGs and MRI template (ICBM152) were co-registered after identifying the anatomical landmarks (left and right pre-auricular points and nasion) using Brainstorm (Tadel *et al.*, 2011). An atlas-based segmentation approach was used to project EEGs onto an anatomical framework consisting of 68 cortical regions identified by means of Desikan-Killiany (Desikan *et al.*, 2006) atlas, see Table S1 (supplementary materials) for more details about the names and abbreviations of these regions. The lead field matrix was then computed for a cortical mesh of 15000 vertices using OpenMEEG (Gramfort *et al.*, 2010).

**Brain networks construction:**

Brain networks were constructed using the 'EEG source connectivity" method (Hassan *et al.*, 2014). It includes two main steps: 1) Reconstruct the temporal dynamics of the cortical sources by solving the inverse problem, and 2) Measure the functional connectivity between the reconstructed time series. Here, we used the weighted minimum norm estimate (wMNE) algorithm as inverse solution (Hamalainen and Ilmoniemi, 1994). The reconstructed regional time series were filtered in different frequency bands [theta (4–8 Hz); alpha1 (8–10 Hz); alpha2 (10–13 Hz); beta (13–30 Hz)]. The functional connectivity was computed, for each frequency band, between the regional time series using the phase locking value (PLV) measure (Lachaux *et al.*, 1999). The PLV ranges between 0 (no phase locking) and 1 (full synchronization).

Using PLV, dynamic functional connectivity matrices were computed for each epoch using a sliding window technique (Kabbara *et al.*, 2017). It consists in moving a time window of certain duration $\delta$ along the time dimension of the epoch, and then PLV is calculated within each window. As recommended in (Lachaux *et al.*, 2000), we chose the smallest window length that is equal to $\frac{6}{\text{central frequency}}$ where 6 is the number of 'cycles' at the given frequency band. In theta band, as the central frequency (Cf) equals to 6 Hz, $\delta$ equals 1s. Likewise, $\delta$ =666 ms in alpha1 band (Cf=9 Hz), 521 ms in alpha2 band (Cf=11.5 Hz), and 279 ms (Cf=21.5 Hz) in beta band. Functional connectivity matrices were represented as graphs (i.e networks) composed of nodes, represented by the 68 ROIs, and edges corresponding to the functional connectivity values computed over the 68 regions, pair-wise.

Considered $\delta$ values yield, for each epoch, to 33 networks in theta band, 66 networks in alpha1 band, 76 networks in alpha2 band and 130 networks in beta band.

**Multi-slice networks modularity:**

The modularity aims at decomposing a network into different communities of high intrinsic connectivity and low extrinsic connectivity (Eickhoff *et al.*, 2005). To describe and quantify the evolution of brain networks as a function of time, we applied the multi-slice modularity (Bassett *et al.*, 2013). In this method, the nodes across network slices (time windows) are linked via a coupling parameter using a quality function given by the following formula:

$$Q_{ml} = \frac{1}{2\mu} \sum_{ijlr} \left\{ \left( A_{ijl} - \gamma_l \frac{k_{il} k_{jl}}{2m_l} \right) \delta_{lr} + \delta_{ij} C_{jlr} \right\} \delta(M_{il}, M_{jr}) \quad (1)$$

Where nodes *i* and *j* are assigned to communities $M_{il}$ and $M_{jl}$ in slice *l*, respectively. $A_{ijl}$ represents the weight of the edge between *i* and *j*. $\gamma_l$ is the structural resolution parameter of slice *l*. $C_{jlr}$ is the connection strength between the node *j* in slice *r* and the node *j* in slice *l*. The structural resolution parameter $\gamma$ and the inter-slice coupling parameter are set to 1. $k_{il}$ is the strength of the node *i* in slice *l*, the $\delta$-function $\delta(x, y)$ is 1 if $x = y$ and 0 otherwise, $m = \frac{1}{2} \sum_{ij} A_{ij}$ and $\mu = \frac{1}{2} \sum_{jr} k_{jr}$.

The multi-slice modularity algorithm was applied with diagonal and ordinal inter-slice couplings. Diagonal and ordinal coupling means that each node is only connected to itself in the adjacent slices. Here, a slice corresponds to a network at a given time period. Hence, the number of slices equals the number of windows at a given frequency band.

To deal with the 'degeneracy' problem, we computed a 68*68 association matrix (Sales-Pardo *et al.*, 2007; Rubinov and Sporns, 2011; Lancichinetti and Fortunato, 2012) where the element $A_{i,j}$ represents the number of times the nodes $i$ and $j$ are assigned to the same module across 200 runs using Louvain algorithm (Blondel *et al.*, 2008). The association matrix was then compared to a null-model generated from 100 random permutations of the original partitions. That is, for each of the 100 partitions, we reassign nodes uniformly at random to the modules present in the partition. This generates a null model matrix whose element $A_{i,j}^{r}$ is the number of times the node $i$ and $j$ are randomly assigned to the same community. To remove randomness, we kept the significant values of the original association matrix by setting any element $A_{i,j}$ whose value is less than the maximum value of the random association matrix to 0 (Bassett *et al.*, 2013). Finally, the thresholded association matrix was re-clustered using Louvain algorithm.

**Network measures**

The topological properties of identified networks were characterized using the following graph measures:

*Average clustering coefficient:* The clustering coefficient of a node represents how close its neighbors tend to cluster together (Watts and Strogatz, 1998). Accordingly, the average clustering coefficient of a network is considered as a direct measure of its segregation (i.e the degree to which a network is organized into local specialized regions) (Bullmore *et al.*, 2009). In brief, the clustering coefficient of a node is defined as the proportion of connections among its neighbors, divided by the number of connections that could possibly exist between them (Watts and Strogatz, 1998).

*Global efficiency:* The global efficiency of a network is the average inverse shortest path length (Latora and Marchiori, 2001). A short path length indicates that, on average, each node can reach other nodes with a path composed of only a few edges (Sporns, 2010). Thus, the global efficiency is one of the most elementary indicators of network's integration (i.e the degree to which a network can share information between distributed regions).

*Recruitment:* The recruitment of a node *i* corresponds to the average probability that the node is in the same module across runs and slices (i.e time windows). It is calculated as follows:

$$Recruitment \ _i^M M_i = \frac{1}{n_M} \sum_{j \in M} A_{i,j} \qquad (2)$$

Where *M* is the module of the node *i*. $n_M$ denotes the number of nodes assigned to the module M. $A_{i,j}$ represents the number of times the nodes *i* and *j* are assigned to the same module across slices and runs. A region with high recruitment value tends to maintain itself in the same community across time (Bassett *et al.*, 2015).

*Integration*: It reflects how modules are interacting with each other. It is computed as the average number of links each node in a given module has with the nodes in the other modules across runs and slices (i.e time windows). It is calculated as follows:

$$Integration \ _i^M M_i = \frac{1}{N - n_M} \sum_{j \notin M} A_{i,j} \qquad (3)$$

Where *M* is the module of the node *i*. *N* denotes the total nodes number, $n_M$ the number of nodes assigned to the module *M*. $A_{i,j}$ represents the number of times the nodes *i* and *j* are assigned to the same module across slices and runs. A region with high integration value tends to be present in communities other than its own across time (Bassett *et al.*, 2015).

## Hubs identification

*Hubness* is a key feature when exploring the brain network architecture due to the high influence of hub nodes on network dynamics and information processing (van den Heuvel and Sporns, 2013). Once modules are identified, the 68 nodes were classified into three main categories (non hubs, provincial hubs and connector hubs) using combination of two measures. The first one is the within-module degree Z defined as:

$$Z_i = \frac{K_i(M_i) - \overline{K(M_i)}}{\sigma_{k(M_i)}} \qquad (4)$$

Where $K_i(M_i)$ is the within-module degree of the node $i$, $\overline{K(M_i)}$ is the mean of within module degree of nodes assigned to the same community as node $i$, and $\sigma_{k(M_i)}$ is the standard deviation. A positive Z value indicates that the node is highly connected to other members of the same community (Guimera, Guimerà and Nunes Amaral, 2005). In our study, a node is considered as hub if the corresponding within module degree is greater than 1.5.

We then focused on classifying hubs into provincial and connector based on a second metric known as participation coefficient (P). This metric characterizes how a node's edges are distributed across modules:

$$P_i = 1 - \sum_{c=1}^{C} \left(\frac{K_i(M)}{K_i}\right)^2 \qquad (5)$$

Where C is the number of modules, $K_i(m)$ is the number of edges between node $i$ and nodes in module M. Based on the criteria proposed by (Guimerà and Nunes Amaral, 2005), a provincial hub having most of its links inside its own module has a $P_i$ value lower than 0.3; while a connector hub has a $P_i$ value greater than 0.3. These values were used in our study.

## Attacks on nodes

Like any other networked system, the brain network may lose some of its effectiveness as a result of an "attack". In particular, attacks on regions playing a key role will lead to significant network disruption. For this reason, we quantified the importance of each node in terms of its attack influence on the global network efficiency. This quantification is usually done using a graph measure known as "vulnerability". It is defined as the reduction in global efficiency of the network when the node and all its edges are removed (Gol'dshtein, Koganov and Surdutovich, 2004). Thus, critical nodes can be identified from high vulnerability values as their attack (i.e node and associated edges removal) leads to significant drop of the whole network efficiency.

## Statistical tests

To quantify the differences between healthy and AD networks in terms of RSNs connectivity, average clustering coefficient, global efficiency, integration/segregation measures and vulnerability, statistical tests were performed. For each subject, we averaged all the metrics values obtained from the different networks among all epochs and time windows for each subject. As data were not normally distributed, we assessed the statistical difference between the two groups using the Mann Whitney U Test also known as Rank-Sum Wilcoxon test (degree of freedom=18).

For hubs identification, each group was considered separately. First, we concatenated the metrics values (participation coefficient and within-module degree Z) from all group subjects, epochs and time windows. Based on the criteria of hubs classification (Guimerà and Nunes Amaral, 2005), each node was assigned to its corresponding category (i.e provincial, connector or non-hub) for each window. Then, the brain regions that are significantly behaving as connector or/and provincial hubs during time were extracted using a chi-squared test (as described in our previous work (Kabbara *et al.*, 2017)). To deal with the family-wise error rate, the statistical tests were corrected for multiple comparisons using Bonferroni method ( $p_{Bonferroni\ adjusted} < \frac{0.05}{N}$ ), with *N* (68) denotes the number of brain regions.

**The parcellation into RSNs**

Each brain region of the Desikan-Killiany atlas was associated to its corresponding RSN based on Shirer et al. (Shirer *et al.*, 2012) in which authors identified fourteen functional networks: anterior salience network, auditory network, basal ganglia network, dorsal default mode network, higher visual network, language network, left executive control network, sensorimotor network, posterior salience network, precunues network, primary visual network, right executive control network, ventral default mode network, and visuospatial network. Here, we focused on five RSNs: the default mode network (DMN) obtained by combining the regions of the dorsal and the ventral default mode network, the salience network (SAN) obtained by associating all the regions in anterior and posterior salience networks, the visual network (VIS) obtained by combining of the higher and primary visual networks. This same parcellation was also used in our previous study (Kabbara *et al.*, 2017).

# Results

**Intrinsic connectivity of RSNs**

First, we were interested in evaluating the differences among the RSNs between healthy controls and AD patients. For this reason, we associated each brain region of the Desikan-Killiany atlas to its corresponding RSN according to (Kabbara *et al.*, 2017). Results in Table 1 show significant decreases in DMN connectivity in AD compared to healthy controls in the theta *(p=0.02, U=15, r=0.51)* and alpha2 *(p=0.031, U=17, r=0.47)* bands. Similarly, reduced visual network connectivity was found in beta band *(p=0.003, U=5, r=0.72)*. Conversely, increased SAN connectivity was observed in the theta band *(p=0.047, U=15.5, r=0.5)*.

**Network integration and segregation**

Here, we explored the difference of brain network dynamics between the two groups in terms of segregation using clustering coefficient and integration using the global efficiency measures. No group difference was observed in alpha1, alpha2 and beta bands. In contrast, in theta band, an increase in clustering coefficient *(p=0.006; U=9, r=0.57)* associated with a decrease in global efficiency *(p=0.03; U=16, r=0.49)* was found in AD networks.

To better explore the difference between the two groups, we clustered the networks into sub-networks (i.e modules or communities) for which the integration and the segregation parameters were extracted. AD networks were characterized by a low inter-modular activity (low integration) and high intra-modular connectivity (high segregation) in theta (Figure 2), alpha1 (Figure S1, supplementary materials), and alpha2 (Figure S2, supplementary materials) bands in contrast with results obtained in beta band (Figure S3, supplementary materials).

| Frequency band | RSN | Healthy | | Alzheimer | | *p*-value |
|---|---|---|---|---|---|---|
| | | Median | SD | Median | SD | |
| Theta | DMN | 0.13 | 0.01 | 0.09 | 0.017 | **0.02*** |
| | DAN | 0.09 | 0.009 | 0.08 | 0.11 | 0.82 |
| | SAN | 0.06 | 0.013 | 0.08 | 0.01 | **0.047*** |
| | AUD | 0.012 | 0.017 | 0.03 | 0.02 | 0.14 |
| | VIS | 0.11 | 0.01 | 0.12 | 0.015 | 0.6 |
| Alpha1 | DMN | 0.117 | 0.018 | 0.116 | 0.014 | 0.68 |
| | DAN | 0.087 | 0.0138 | 0.0975 | 0.008 | 0.21 |
| | SAN | 0.057 | 0.0101 | 0.07 | 0.014 | 0.17 |
| | AUD | 0.024 | 0.011 | 0.023 | 0.018 | 0.75 |
| | VIS | 0.15 | 0.017 | 0.13 | 0.018 | 0.11 |
| Alpha2 | DMN | 0.12 | 0.008 | 0.11 | 0.005 | **0.031*** |
| | DAN | 0.095 | 0.0037 | 0.091 | 0.009 | 0.35 |
| | SAN | 0.076 | 0.01 | 0.071 | 0.012 | 0.4 |
| | AUD | 0.036 | 0.006 | 0.033 | 0.012 | 0.4 |
| | VIS | 0.1311 | 0.1084 | 0.12 | 0.019 | 0.35 |
| Beta | DMN | 0.12 | 0.012 | 0.12 | 0.007 | 0.3 |
| | DAN | 0.091 | 0.044 | 0.086 | 0.011 | 0.25 |
| | SAN | 0.069 | 0.0075 | 0.074 | 0.0012 | 0.16 |
| | AUD | 0.033 | 0.013 | 0.02 | 0.012 | 0.09 |
| | VIS | 0.134 | 0.014 | 0.108 | 0.012 | **0.003*** |

**Table 1. Differences among RSNs connectivity between healthy and AD networks in the different frequency bands. Abbreviations: Default mode network=DMN, Dorsal attention network=DAN, Salience attention network=SAN, Auditory network=AUD, Visual network=VIS. * denotes for significant effects ( *p*<0.05).**

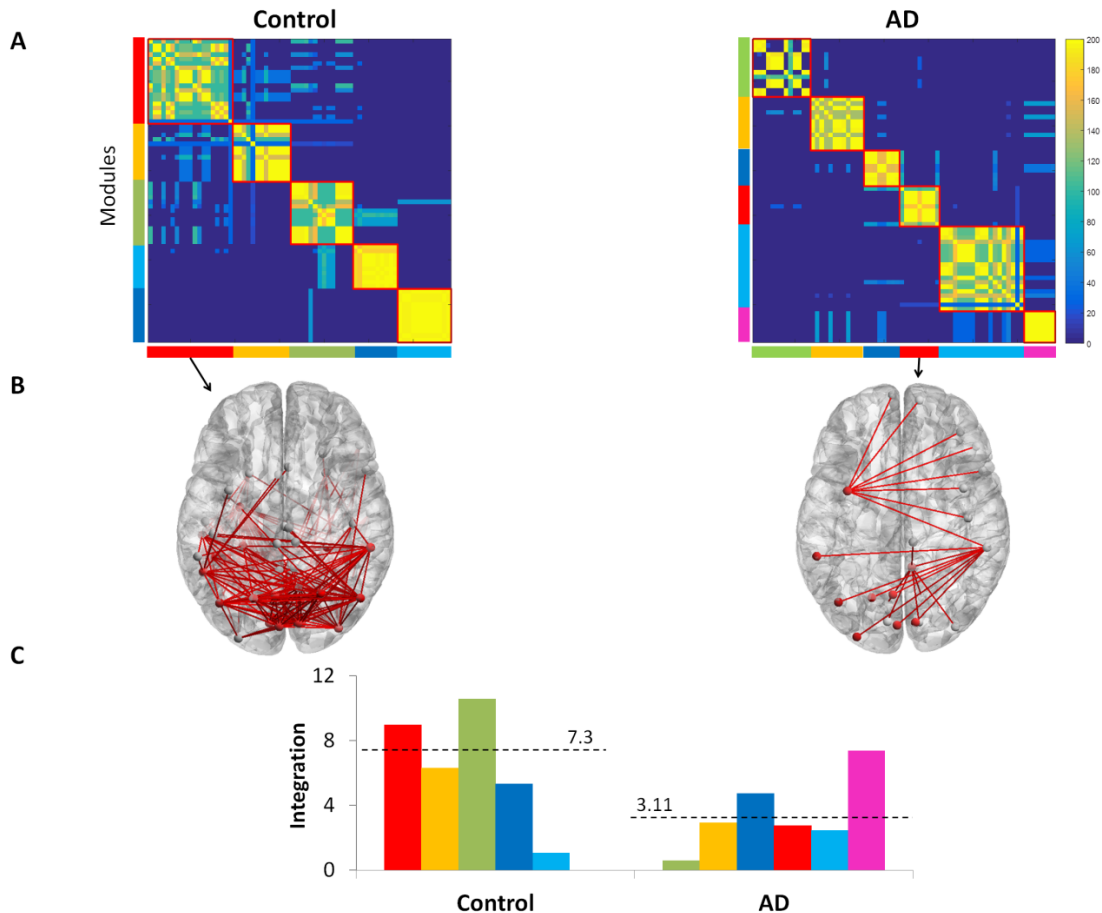

**Figure 2.** A) The association matrices obtained by the multi-slice modularity method for healthy controls and AD patients. B) A typical example showing the difference between the inter-modular interactions obtained for the red module in both groups. C) The bar plots show the integration values of each group's modules. The dotted line presents the average integration value across modules.

## Hubs identification

The cortical distributions of connector and provincial hubs identified in healthy subjects and AD patients are illustrated in Figure 3. A loss in connector hubs number was observed in AD networks, while the number of provincial hubs was found to increase compared to healthy networks. Specifically, only the left middle orbito-frontal region was conserved in AD network as a connector hub, whereas the right middle orbito-frontal, the left rostral anterior cingulate, the right transverse temporal, the left posterior cingulate, the right posterior cingulate, the right isthmus cingulate and the left precunues regions were present in healthy networks. In contrast, the left middle orbito-frontal, the right middle orbito-frontal and the right insula appeared as provincial hubs in AD networks.

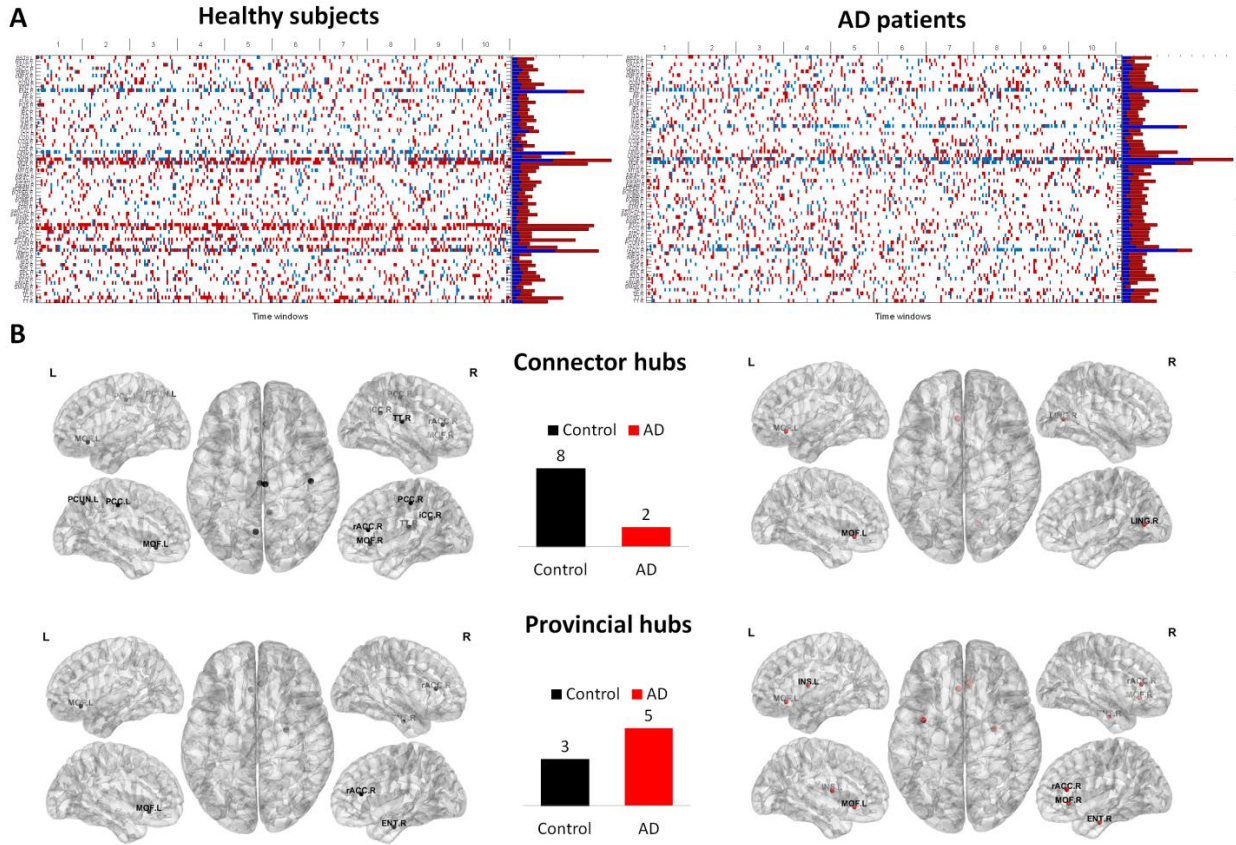

**Figure 3. A) Variations of the node type (provincial vs. connector) across time and subjects for the 68 brain regions in both groups. Bar plots represent the number of times a node is considered as provincial hub (blue color) and as connector hub (red color). B) The spatial distributions of significant provincial hubs, and significant connector hubs in both groups ( $p_{Bonferroni\ adjusted} < \frac{0.05}{68}$ ). Bar plots illustrate the difference in the number of connector and provincial hubs between the two groups.**

We then investigated the influence of each node's removal on the global efficiency of the networks using the vulnerability metric. Results are shown in Figure 4. We realized that 11 brain regions were more vulnerable in healthy networks versus AD networks (*p*<0.05). However, only the right middle orbito-frontal and the left lateral orbito-frontal regions have resisted the Bonferroni correction ( $p_{Bonferroni\ adjusted} < \frac{0.05}{68}$ ).While the 11 nodes are distributed across several RSNs, the majority of these regions corresponds to DMN (6/11) including mainly the isthmus cingulate, the middle orbito-frontal and the rostral cingulate.

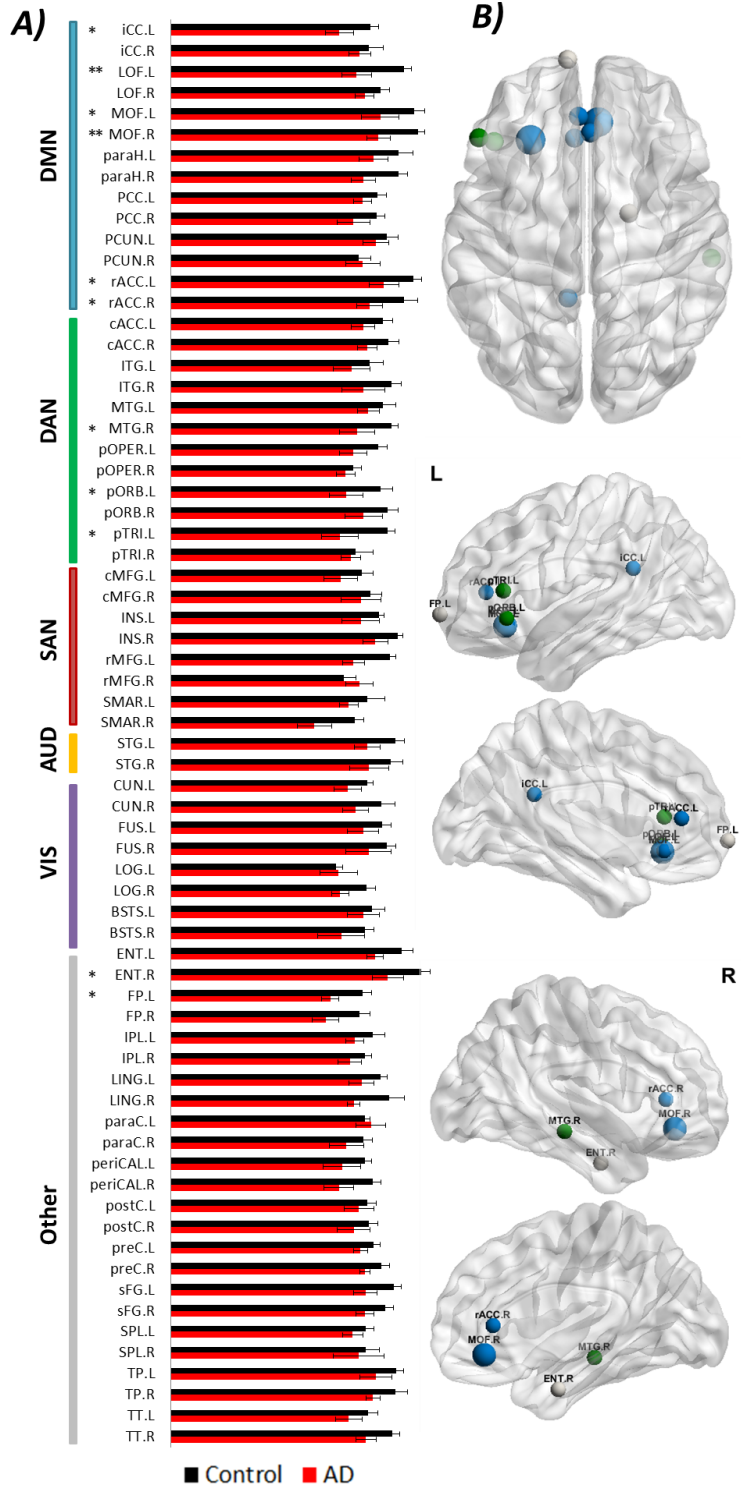

**Figure 4. Difference between healthy subjects and AD patients in term of node vulnerability. A)** Distribution of vulnerability values for the 68 ROIs in healthy control networks (black color) and in AD networks (red color). A node is marked with * if it shows significant difference between groups ($p<0.05$, uncorrected) and with ** if it shows significant difference after correction for multiple comparisons. **B)** Cortical distribution of the 11 significant nodes. The node color corresponds to the matching RSN (see Table 1 for ROI names and abbreviations). The nodes with larger size are those who resisted the multiple comparison adjustment.

**Correlation between network measures and cognitive scores**

To assess the relationships between functional connectivity and the AD patient's cognitive impairment, we have estimated the correlation between the cognitive score (MMSE) and the network measures (clustering coefficient, global efficiency and vulnerability). A negative correlation between the average clustering coefficient and MMSE score ($\rho= -0.95$; $p<0.001$) was found, while a positive correlation between the network global efficiency and MMSE score ($\rho= 0.94$; $p<0.001$) was obtained (Figure 5). Concerning the vulnerability, we focused on the two nodes that showed statistical difference between groups. Figure 5 shows that the MMSE score correlates positively with the left lateral orbito-frontal region ($\rho= 0.84$; $p=0.002$), and the right middle orbito-frontal region ($\rho= 0.87$; $p=0.001$).

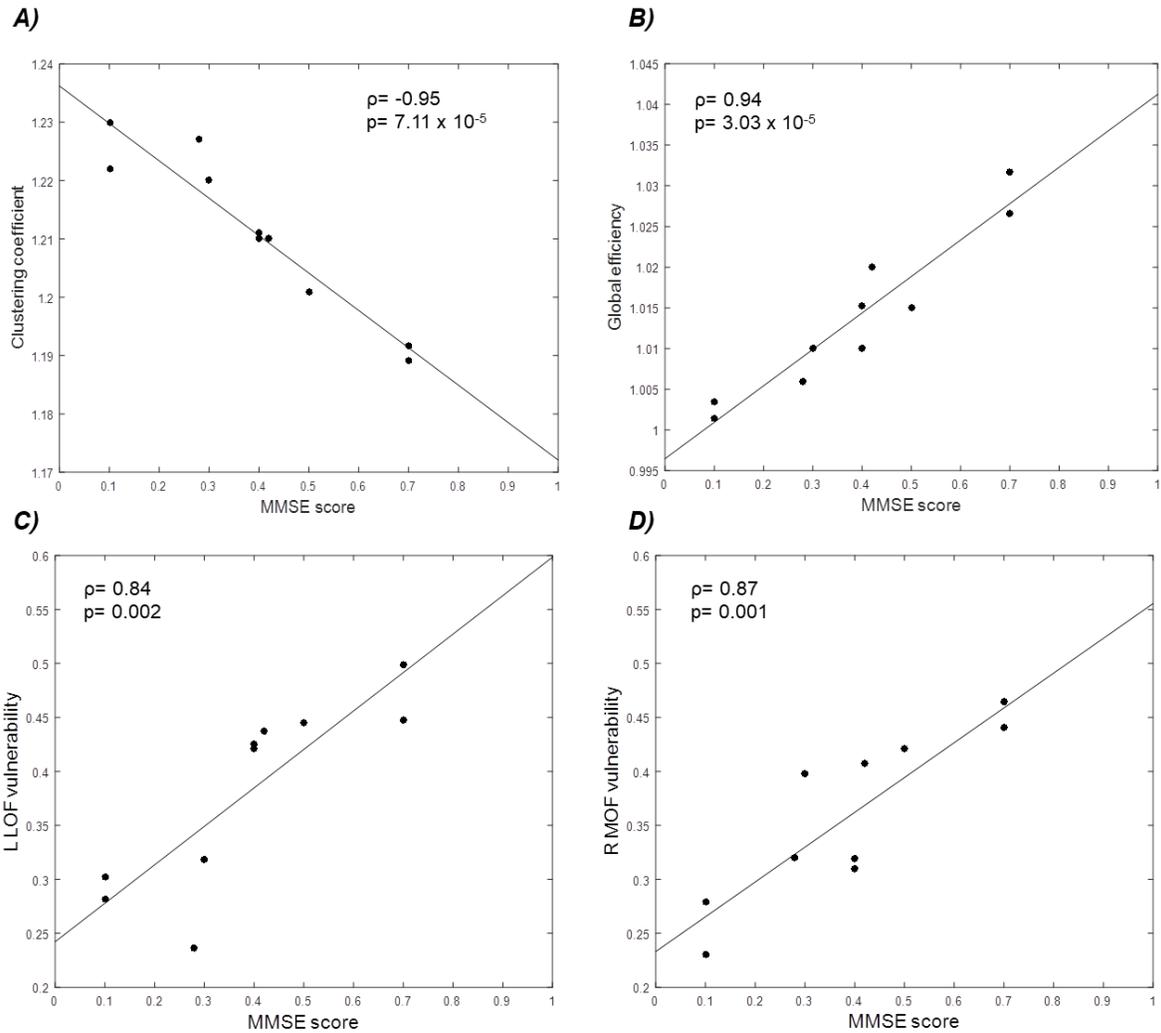

Figure 5. Correlation between the cognitive score (MMSE) and the graph measures for AD patients. A) Clustering coefficient and B) Global efficiency. C) Vulnerability of the left lateral orbito-frontal region. D) Vulnerability of the right middle orbito-frontal region.

# Discussion

The main objective in this study is to explore the dynamic topological properties of AD networks compared to healthy controls. Particularly, we focused on examining the shifting balance between brain network integration and segregation in Alzheimer's disease. For this end, resting state EEG signals were recorded from 20 participants (10 AD patients and 10 controls). The cortical functional networks were reconstructed from scalp signals using the EEG source connectivity method. A sliding window approach was used to track the dynamics of networks. To examine the differences between the two groups (AD vs. controls), several network measures were extracted. The measures used to quantify the integration of networks are: the network global efficiency, the inter-modular connections and the connector hubs. To quantify segregation we extracted the clustering coefficient, the intra-modular connections and the provincial hubs. The nodes resilience against attacks was also analyzed in order to identify the main brain regions potentially affected by AD. Interestingly, a general trend is that all metrics showed that AD networks tend to have improved segregation (higher local information processing) and reduced integration (lower global information processing). Results also showed a significant correlation between patients' cognitive performance (as measured by the MMSE score) and network measures. Results are discussed in detail hereafter.

## AD networks: high segregation and low integration

Results indicated that AD networks are characterized by lower integration (revealed by a decrease in the network global efficiency, the number of connector hubs and the integration measure), and higher segregation (revealed by an increase in clustering coefficient, in the number of provincial hubs and in the recruitment measure) compared to healthy control networks. One possible interpretation of the increased local connectivity is a possible compensatory mechanism that is triggered by the dysfunctional integration in the AD brain networks (Afshari and Jalili, 2016). These findings are in line with studies that revealed decrease in the network global efficiency (Stam *et al.*, 2009; Lo *et al.*, 2010; Douw *et al.*, 2011; Stam and van Straaten, 2012; Zhao *et al.*, 2012; Tijms, Wink, *et al.*, 2013; Afshari and Jalili, 2016) and the participation coefficient (De Haan *et al.*, 2012) in AD networks. In line with these studies, Debeuck and coll. (Delbeuck *et al.*, 2003) studied the McGurk effect in AD and reported that that the integration between auditory and visual speech information was disrupted .The increased segregation observed in AD was reported using the local efficiency and the clustering coefficient (Zhao *et al.*, 2012; Afshari and Jalili, 2016). More importantly, , and in line with our findings, a longitudinal EEG study reported reduced global efficiency and increased clustering coefficient during AD progression (Morabito *et al.*, 2015).

**EEG Frequency bands**

EEG is increasingly used to detect cognitive deficits in neurodegenerative disorders. One of the main and consistent findings is the shift to lower frequencies in Alzheimer's disease, using resting-state recordings (Bennys et al., 2001). A slowing of EEGs in the theta power was also observed in Alzheimer's disease at early stage of the disease (Benz et al., 2014). Several previous studies have confirmed the importance of the theta band with regards to cognition, see (Klimesch, 1999; Axmacher *et al.*, 2006) for two reviews. Moreover, the importance of theta activity in controlling the working memory processes was widely reported (Sarnthein *et al.*, 1998; Klimesch, 1999; Stam, 2000; Stam and Van Dijk, 2002; Sauseng *et al.*, 2010). Our findings are in accordance with these studies. A potential interpretation of these findings is that disruption of lower frequencies such as theta rhythms is due to degeneration processes in the attentional system (Hassan et al., 2017; Klimesch, 1999).

Compared to other frequency bands, here we found significant differences in theta band network characteristics in AD networks, namely, lower integration (low global efficiency), higher segregation (high recruitment and average clustering) , a lower number of hubs, a lower effect of nodes' removal and a disrupted function of DMN. Abnormal EEG correlations in parietal and frontal regions within alpha and theta bands were reported in early AD stage (Montez *et al.*, 2009). Using brain network analysis, several previous studies have observed alterations in the lower frequency bands in patients with dementia. These findings revealed loss in hubs, disruption in functional connectivity (Bosboom *et al.*, 2009), reduction in network efficiency (van Dellen *et al.*, 2015) and a decrease in local integration (Utianski *et al.*, 2016) in the alpha2 band.

Results also depict an opposite influence of the lower frequency bands (theta, alpha1, alpha2) on the balance of integration/segregation compared to the higher frequency band (beta). A possible explanation is the complementary role of frequencies in conducting long/short range connections. In fact, while integrated information is mediated by low frequency bands, local information processing is mediated by high frequency bands (Von Stein and Sarnthein, 2000; Buzsáki and Draguhn, 2004; Schroeder and Lakatos, 2009; Canolty and Knight, 2010; Siegel, Donner and Engel, 2012).

**Altered brain networks/regions in Alzheimer's disease**

On the one side, the detection of nodal changes can reveal important insights about which brain regions are severely altered by the disease. Our results show a change in hub properties for R MOF, L rACC, R TT, L/R pCC and L pCUN (see Table 1 for abbreviations). We also hypothesized that the removal of an important brain region will affect the information processing in the whole network, while an attack to a less critical region will have a smaller influence on the global network efficiency. We found 13 brain regions that have more importance in healthy network than in AD networks. One can realize that some of the affected hubs (Figure 3) coincide with the 13 nodes (rACC, MOF, pCUN, TT). These affected nodes were also reported in (Sorg *et al.*, 2007; Bai *et al.*, 2009; Buckner *et al.*, 2009; Mormino *et al.*, 2011; De Haan *et al.*, 2012;

Vemuri, Jones and Jack Jr., 2012; Tijms, Möller, *et al.*, 2013). Other studies also reported that amyloid decomposition in AD coincide with hubs location (Buckner *et al.*, 2009).

On the other side, alterations in the default mode network (DMN) connectivity in AD patients were reported in several studies (Li *et al.*, 2002; Greicius *et al.*, 2004; Wang *et al.*, 2006; Sorg *et al.*, 2007; Hedden *et al.*, 2009; Sheline *et al.*, 2010; Drzezga *et al.*, 2011; Mormino *et al.*, 2011; Vemuri, Jones and Jack Jr., 2012). Our results showed that the majority of the affected nodes in terms of vulnerability and hub dys-functionality are associated to the DMN. The disruption of DMN was also demonstrated by its reduced intrinsic connectivity as reported in Table 1. The increased connectivity of DAN and SAN shown in Table 1 may be interpreted as a compensatory mechanism due to the DMN alteration (Bai *et al.*, 2011; Damoiseaux *et al.*, 2012).

**Correlation between network measures and AD patient's cognitive scores**

Single-subject analyses showed significant correlation between the MMSE score (used here to provide an overall measure of cognitive impairment) and network global efficiency, average clustering coefficient and vulnerability. Although the MMSE test has received good acceptance as a diagnostic test in the clinical and research community (Nieuwenhuis-Mark, 2010), it is recommended not to be used as a stand-alone single administration test (Arevalo-Rodriguez *et al.*, 2015). Previous studies have shown that age, education and socio-cultural variables affect the effectiveness of MMSE to detect cognitive impairment (Bleecker *et al.*, 1988; Brayne and Calloway, 1990; Crum, 1993). Hence, the demand is high for other tests that provide higher detection accuracy (Carnero-Pardo *et al.*, 2011, 2014), as well as more specific scores (semantic, memory related… etc.). In addition, the use of cognitive tasks that stimulate the affected networks in the case of AD (the memory network for instance) may improve the correlations with network based metrics. It is worth noting that the MMSE is not the unique test for AD diagnosis. It is currently used within a set of other tests including clinical examination (reflexes, muscle tone, balance) and brain imaging (such MRI and CT scan) aimed to pinpoint visible abnormalities related to conditions other than AD (stroke, trauma.. etc.). However, when MRI is negative (no visible anatomical damages), the screening of cognitive performance using clinical tests such as MMSE (or other specific cognitive scores) is mandatory. The proposed network-based metrics provides additional quantitative indications potentially useful for neurologists to complement diagnosis based on neuropsychological tests.

## Limitations

First, one of the main limitations of this study is the relatively low number of patients. Our intent was to show the difference between two groups: totally normal (control group) and AD patients with 'severe' cognitive impairment. Nevertheless, we are aware that the AD is very heterogeneous and may have different stages including patients with moderate or mild cognitive impairment. Detecting these 'early' cognitive deficits is on the major challenges in AD and will certainly be the subject of future investigation. These investigations should be performed on larger cohorts of patients in different AD stages, using other experimental paradigms and additional cognitive scores, in order to be able to generalize the conclusions of the reported analysis.

Second, the EEG source connectivity was applied here to 32 scalp EEG channels. This method has previously proved its robustness in exploring resting-state topology using dense-EEG (>128 electrodes) (Kabbara *et al.*, 2016, 2017; Hassan *et al.*, 2017). As reported in (Hassan *et al.*, 2014), the use of a smaller number of electrodes (in the context of cognitive task) will result in a reduction in the accuracy of the obtained results. Nevertheless, several studies showed the possible extraction of useful information using low number of electrodes (19, 32, 64) (Canuet *et al.*, 2012; Vecchio *et al.*, 2014, 2017; Hata *et al.*, 2016). This can be explained by the facts that these studies (as the presented study) focus on the investigation of 'large-scale' networks to compare two groups with the same conditions. In addition, we conjecture that a compromise between the number of channels and the number of ROIs should be necessarily respected. Our very recent findings showed that a high number of electrodes (>32) is mandatory in the case of applications that require higher "granularity", i.e. spatial precision and accurate characterization of the network local properties, such as the identification of epileptogenic networks (*unpublished data*).

Third, it is important to keep in mind that measuring the functional connectivity is generally corrupted by the volume conduction problem (Schoffelen and Gross, 2009). While the effects of this problem are reduced by the analysis of connectivity at source level, some "mixing effects" remain (Brookes, Woolrich and Price, 2014). At the source level, few strategies have been suggested (Brookes, Woolrich and Barnes, 2012; Colclough *et al.*, 2015). The proposed approaches are all based on ignoring zero-lag interactions among signals, by supposing that their contributions are only due to the source leakage. Although these approaches have some advantages, they may also remove true communications that occur at zero lag (Finger *et al.*, 2016). In our study, we used the phase locking value measure. In a previous study, we showed that the metrics extracted from the networks constructed using PLV (including the within-module degree, clustering coefficient, betweenness centrality and the participation coefficient) were not affected by the spurious short connections (Kabbara *et al.*, 2017). Nevertheless, we believe that further methodological efforts are needed to completely solve the spatial leakage problem.

Fourth, a proportional threshold of 10% was used to remove the spurious connections from the connectivity matrices. Here, we preferred using a proportional threshold to absolute threshold to ensure equal density between groups, as recommended by (van den Heuvel *et al.*, 2017). Moreover, Garisson et al.(Garrison *et al.*, 2015) showed that network measures are stable across proportional thresholds, in contrast to absolute thresholds. A variety of thresholding methods are available, but no method is free of bias. It is then recommended to perform studies across different values of thresholds (in addition to the use of alternative strategies) to ensure that the obtained findings are robust to this methodological factor.

Fifth, the choice of the inverse solution/connectivity combination was supported by two comparative studies using simulated data from a biophysical/physiological model (Hassan *et al.*, 2016) and real data recorded during a cognitive task (Hassan *et al.*, 2014). In both analyses, the combination that showed the highest similarity between reference (ground truth) and estimated networks was the wMNE/PLV, used in the present paper. Nevertheless, other combinations or strategies that showed accurate construction of cortical networks from sensor level recordings could be also investigated and compared such as the use of beamforming combined with amplitude correlation between band-limited power envelops as reported in several studies (Brookes *et al.*, 2011; Brookes, Woolrich and Barnes, 2012; Colclough *et al.*, 2015, 2016; O'Neill *et al.*, 2016).

# Conclusions

We reported a study using EEG connectivity at the source level in AD patients and healthy controls. We showed that AD networks are characterized by a reduction in their global performance (integration) associated with an enhancement in their local performance (segregation). We also showed that these network topologies are correlated with the patient's cognitive scores. We speculate that our findings, when validated on larger cohort, could contribute to the development of EEG-based tests that could consolidate results of currently-used neurophysiological tests.

# Acknowledgements

This study was funded by the National Council for Scientific Research (CNRS) in Lebanon. The work has also received a French government support granted to the CominLabs excellence laboratory and managed by the National Research Agency in the "Investing for the Future" program under reference ANR-10-LABX-07-01. It was also financed by Azm Center for research in biotechnology and its applications.

# References

Afshari, S. and Jalili, M. (2016) 'Directed Functional Networks in Alzheimer's Disease: Disruption of Global and Local Connectivity Measures', *IEEE Journal of Biomedical and Health Informatics*, 2194(c), pp. 1–1. doi: 10.1109/JBHI.2016.2578954.
Arevalo-Rodriguez, I. *et al.* (2015) 'Mini-Mental State Examination (MMSE) for the detection of Alzheimer's disease and other dementias in people with mild cognitive impairment (MCI)', *The Cochrane database of systematic reviews*, p. CD010783. doi: 10.1002/14651858.CD010783.pub2.
Axmacher, N. *et al.* (2006) 'Memory formation by neuronal synchronization', *Brain Research Reviews*, pp. 170–182. doi: 10.1016/j.brainresrev.2006.01.007.
Bai, F. *et al.* (2009) 'Abnormal resting-state functional connectivity of posterior cingulate cortex in amnestic type mild cognitive impairment', *Brain Research*, 1302, pp. 167–174. doi: 10.1016/j.brainres.2009.09.028.
Bai, F. *et al.* (2011) 'Specifically progressive deficits of brain functional marker in amnestic type mild cognitive impairment', *PLoS ONE*, 6(9). doi: 10.1371/journal.pone.0024271.
Bassett, D. S. *et al.* (2013) 'Robust detection of dynamic community structure in networks', *Chaos*, 23(1). doi: 10.1063/1.4790830.
Bassett, D. S. *et al.* (2015) 'Learning-Induced Autonomy of Sensorimotor Systems', *Nature neuroscience*, 18(5), pp. 744–751. doi: 10.1038/nn.3993.
Bennys, K., Rondouin, G., Vergnes, C., and Touchon, J. (2001). Diagnostic value of quantitative EEG in Alzheimer's disease. Neurophysiologie Clinique/Clinical Neurophysiology 31, 153-160.

Benz, N., Hatz, F., Bousleiman, H., Ehrensperger, M.M., Gschwandtner, U., Hardmeier, M., Ruegg, S., Schindler, C., Zimmermann, R., and Monsch, A.U. (2014). Slowing of EEG background activity in Parkinson's and Alzheimer's disease with early cognitive dysfunction. Frontiers in aging neuroscience 6.

Bleecker, M. L. *et al.* (1988) 'Age-specific norms for the Mini-Mental State Exam.', *Neurology*, 38(10), pp. 1565–8. doi: 10.1212/WNL.38.10.1565.

Blondel, V. D. *et al.* (2008) 'Fast unfolding of communities in large networks', *Journal of Statistical Mechanics: Theory and Experiment*, 10008(10), p. 6. doi: 10.1088/1742-5468/2008/10/P10008.

Bosboom, J. L. W. *et al.* (2009) 'MEG resting state functional connectivity in Parkinson's disease related dementia', *Journal of Neural Transmission*, 116(2), pp. 193–202. doi: 10.1007/s00702-008-0132-6.

Brayne, C. and Calloway, P. (1990) 'The association of education and socioeconomic status with the Mini Mental State Examination and the clinical diagnosis of dementia in elderly people.', *Age and ageing*, 19, pp. 91–96.

Brookes, M. J. *et al.* (2011) 'Measuring functional connectivity using MEG: Methodology and comparison with fcMRI', *NeuroImage*, 56, pp. 1082–1104. doi: 10.1016/j.neuroimage.2011.02.054.

Brookes, M. J., Woolrich, M. W. and Barnes, G. R. (2012) 'Measuring functional connectivity in MEG: A multivariate approach insensitive to linear source leakage', *NeuroImage*, 63(2), pp. 910–920. doi: 10.1016/j.neuroimage.2012.03.048.

Brookes, M. J., Woolrich, M. W. and Price, D. (2014) 'An Introduction to MEG connectivity measurements', in *Magnetoencephalography: From Signals to Dynamic Cortical Networks*, pp. 321–358. doi: 10.1007/978-3-642-33045-2_16.

Buckner, R. L. *et al.* (2009) 'Cortical hubs revealed by intrinsic functional connectivity: mapping, assessment of stability, and relation to Alzheimer's disease', *J Neurosci*, 29(6), pp. 1860–1873. doi: 10.1523/JNEUROSCI.5062-08.2009.

Bullmore, E. *et al.* (2009) 'Complex brain networks: graph theoretical analysis of structural and functional systems', *Nat Rev Neurosci*, 10(3), pp. 186–198. doi: 10.1038/nrn2575.

Buzsáki, G. and Draguhn, A. (2004) 'Neuronal oscillations in cortical networks.', *Science (New York, N.Y.)*, 304(5679), pp. 1926–1929. doi: 10.1126/science.1099745.

Canolty, R. T. and Knight, R. T. (2010) 'The functional role of cross-frequency coupling', *Trends in Cognitive Sciences*, pp. 506–515. doi: 10.1016/j.tics.2010.09.001.

Canuet, L. *et al.* (2012) 'Resting-State Network Disruption and APOE Genotype in Alzheimer's Disease: A lagged Functional Connectivity Study', *PLoS ONE*, 7(9). doi: 10.1371/journal.pone.0046289.

Carnero-Pardo, C. *et al.* (2011) 'Diagnostic accuracy, effectiveness and cost for cognitive impairment and dementia screening of three short cognitive tests applicable to illiterates', *PLoS ONE*, 6(11). doi: 10.1371/journal.pone.0027069.

Carnero-Pardo, C. *et al.* (2014) 'A systematic review and meta-analysis of the diagnostic accuracy of the Phototest for cognitive impairment and dementia', *Dementia & Neuropsychologia*, 8(2), pp. 141–147. doi: 10.1590/S1980-57642014DN82000009.

Colclough, G. L. *et al.* (2015) 'A symmetric multivariate leakage correction for MEG connectomes', *NeuroImage*, 117, pp. 439–448. doi: 10.1016/j.neuroimage.2015.03.071.

Colclough, G. L. *et al.* (2016) 'How reliable are MEG resting-state connectivity metrics?', *NeuroImage*. The Authors, 138, pp. 284–293. doi: 10.1016/j.neuroimage.2016.05.070.

Crum, R. M. (1993) 'Population-based norms for the Mini-Mental State Examination by age and educational level', *JAMA: The Journal of the American Medical Association*, 269(18), pp. 2386–2391. doi: 10.1001/jama.269.18.2386.

Damoiseaux, J. S. *et al.* (2012) 'Functional connectivity tracks clinical deterioration in Alzheimer's disease', *Neurobiology of Aging*, 33(4). doi: 10.1016/j.neurobiolaging.2011.06.024.

Delbeuck, X. *et al.* (2003) 'Alzheimer'Disease as a Disconnection Syndrome?', *Neuropsychology review*, 13(2), pp. 79–92. doi: 10.1023/A:1023832305702.

van Dellen, E. *et al.* (2015) 'Loss of EEG Network Efficiency Is Related to Cognitive Impairment in Dementia With Lewy Bodies.', *Movement disorders : official journal of the Movement Disorder Society*, 30(13), pp. 1785–1793. doi: 10.1002/mds.26309.

Delorme, A. and Makeig, S. (2004) 'EEGLAB: An open source toolbox for analysis of single-trial EEG dynamics including independent component analysis', *Journal of Neuroscience Methods*, 134, pp. 9–21. doi: 10.1016/j.jneumeth.2003.10.009.

Desikan, R. S. *et al.* (2006) 'An automated labeling system for subdividing the human cerebral cortex on MRI scans into gyral based regions of interest', *NeuroImage*, 31, pp. 968–980. doi: 10.1016/j.neuroimage.2006.01.021.

Douw, L. *et al.* (2011) 'Cognition is related to resting-state small-world network topology: An magnetoencephalographic study', *Neuroscience*, 175, pp. 169–177. doi: 10.1016/j.neuroscience.2010.11.039.

Drzezga, A. *et al.* (2011) 'Neuronal dysfunction and disconnection of cortical hubs in non-demented subjects with elevated amyloid burden', *Brain*, 134(6), pp. 1635–1646. doi: 10.1093/brain/awr066.

Eickhoff, S. B. *et al.* (2005) 'A new SPM toolbox for combining probabilistic cytoarchitectonic maps and functional imaging data', *NeuroImage*, 25(4), pp. 1325–1335. doi: 10.1016/j.neuroimage.2004.12.034.

Engels, M. M. A. *et al.* (2017) 'Alzheimer's disease: The state of the art in resting-state magnetoencephalography', *Clinical Neurophysiology*, 128(8), pp. 1426–1437. doi: 10.1016/j.clinph.2017.05.012.

Finger, H. *et al.* (2016) 'Modeling of Large-Scale Functional Brain Networks Based on Structural Connectivity from DTI: Comparison with EEG Derived Phase Coupling Networks and Evaluation of Alternative Methods along the Modeling Path', *PLoS Computational Biology*, 12(8). doi: 10.1371/journal.pcbi.1005025.

Folstein, M. F., Folstein, S. E. and McHugh, P. R. (1975) '"Mini-mental state". A practical method for grading the cognitive state of patients for the clinician', *Journal of Psychiatric Research*, 12(3), pp. 189–198. doi: 10.1016/0022-3956(75)90026-6.

Garrison, K. A. *et al.* (2015) 'The (in)stability of functional brain network measures across thresholds', *NeuroImage*, 118, pp. 651–661. doi: 10.1016/j.neuroimage.2015.05.046.

Gol'dshtein, V., Koganov, G. A. and Surdutovich, G. I. (2004) 'Vulnerability and Hierarchy of Complex Networks', *Physics*, 16, p. 4.

Gramfort, A. *et al.* (2010) 'OpenMEEG: opensource software for quasistatic bioelectromagnetics.', *Biomedical engineering online*, 9, p. 45. doi: 10.1186/1475-925X-9-45.

Greicius, M. D. *et al.* (2004) 'Default-mode network activity distinguishes Alzheimer's disease from healthy aging: evidence from functional MRI.', *Proceedings of the National Academy of Sciences of the United States of America*, 101(13), pp. 4637–42. doi: 10.1073/pnas.0308627101.

Guimera, R., Guimerà, R. and Nunes Amaral, L. a (2005) 'Functional cartography of complex metabolic networks.', *Nature*, 433(7028), pp. 895–900. doi: 10.1038/nature03288.

Guimerà, R. and Nunes Amaral, L. A. (2005) 'Functional cartography of complex metabolic networks', *Nature*, 433(7028), pp. 895–900. doi: 10.1038/nature03288.

De Haan, W. *et al.* (2012) 'Disrupted modular brain dynamics reflect cognitive dysfunction in Alzheimer's disease', *NeuroImage*, 59(4), pp. 3085–3093. doi: 10.1016/j.neuroimage.2011.11.055.

Hamalainen, M. S. and Ilmoniemi, R. J. (1994) 'Interpreting magnetic fields of the brain: minimum norm estimates', *Medical & Biological Engineering & Computing*, 32(1), pp. 35–42. doi: 10.1007/BF02512476.

Hassan, M. *et al.* (2014) 'EEG source connectivity analysis: From dense array recordings to brain networks', *PLoS ONE*, 9. doi: 10.1371/journal.pone.0105041.

Hassan, M. *et al.* (2015) 'Dynamic reorganization of functional brain networks during picture naming', *Cortex*, 73, pp. 276–288. doi: 10.1016/j.cortex.2015.08.019.

Hassan, M. *et al.* (2016) 'Identification of Interictal Epileptic Networks from Dense-EEG', *Brain Topography*, pp. 1–17. doi: 10.1007/s10548-016-0517-z.

Hassan, M. *et al.* (2017) 'Functional connectivity disruptions correlate with cognitive phenotypes in Parkinson's disease', *NeuroImage: Clinical*, 14, pp. 591–601. doi: 10.1016/j.nicl.2017.03.002.

Hata, M. *et al.* (2016) 'Functional connectivity assessed by resting state EEG correlates with cognitive decline of Alzheimer's disease - An eLORETA study', *Clinical Neurophysiology*, 127(2), pp. 1269–1278. doi: 10.1016/j.clinph.2015.10.030.

Hedden, T. *et al.* (2009) 'Disruption of functional connectivity in clinically normal older adults harboring amyloid burden.', *The Journal of neuroscience : the official journal of the Society for Neuroscience*, 29(40), pp. 12686–94. doi: 10.1523/JNEUROSCI.3189-09.2009.

van den Heuvel, M. P. *et al.* (2017) 'Proportional thresholding in resting-state fMRI functional connectivity networks and consequences for patient-control connectome studies: Issues and recommendations', *NeuroImage*, 152, pp. 437–449. doi: 10.1016/j.neuroimage.2017.02.005.

van den Heuvel, M. P. and Sporns, O. (2013) 'Network hubs in the human brain', *Trends in Cognitive Sciences*, pp. 683–696. doi: 10.1016/j.tics.2013.09.012.

Hipp, J. F. *et al.* (2012) 'Large-scale cortical correlation structure of spontaneous oscillatory activity', *Nature Neuroscience*, 15(6), pp. 884–890. doi: 10.1038/nn.3101.

Hughes, S. W. *et al.* (2004) 'Synchronized oscillations at α and θ frequencies in the lateral geniculate nucleus', *Neuron*, 42(2), pp. 253–268. doi: 10.1016/S0896-6273(04)00191-6.

Hughes, S. W. and Crunelli, V. (2005) 'Thalamic Mechanisms of EEG Alpha Rhythms and Their Pathological Implications', *The Neuroscientist*, 11(4), pp. 357–372. doi: 10.1177/1073858405277450.

Ismail, Z., Rajji, T. K. and Shulman, K. I. (2010) 'Brief cognitive screening instruments: An update', *International Journal of Geriatric Psychiatry*, pp. 111–120. doi: 10.1002/gps.2306.

Kabbara, A. *et al.* (2016) 'Graph analysis of spontaneous brain network using EEG source connectivity', *arXiv preprint arXiv:1607.00952*.

Kabbara, A. *et al.* (2017) 'The dynamic functional core network of the human brain at rest'. Springer US, (August 2016), pp. 1–16. doi: 10.1038/s41598-017-03420-6.

Klem, G. *et al.* (1958) 'The ten-twenty electrode system of the International Federation', *Electroencephalography and Clinical Neurophysiology*, 10(2), pp. 371–375. doi: 10.1016/0013-4694(58)90053-1.

Klimesch, W. (1999) 'EEG alpha and theta oscillations reflect cognitive and memory performance: A review and analysis', *Brain Research Reviews*, pp. 169–195. doi: 10.1016/S0165-0173(98)00056-3.

Korjus, K. *et al.* (2015) 'Personality cannot be predicted from the power of resting state EEG', *Frontiers in Human Neuroscience*, 9. doi: 10.3389/fnhum.2015.00063.

Lachaux, J.-P. *et al.* (2000) 'Studying single-trials of phase synchronous activity in the brain', *International Journal of Bifurcation and Chaos*, 10(10), pp. 2429–39. doi: 10.1142/S0218127400001560.

Lachaux, J. P. *et al.* (1999) 'Measuring phase synchrony in brain signals', *Human Brain Mapping*, 8, pp. 194–208. doi: 10.1002/(SICI)1097-0193(1999)8:4<194::AID-HBM4>3.0.CO;2-C.

Lancichinetti, A. and Fortunato, S. (2012) 'Consensus clustering in complex networks.', *Scientific reports*, 2, p. 336. doi: 10.1038/srep00336.

Latora, V. and Marchiori, M. (2001) 'Efficient Behavior of Small World Networks', *Physical Review Letters*, 87, p. 198701. doi: 10.1103/PhysRevLett.87.198701.

Li, F. *et al.* (2015) 'Relationships between the resting-state network and the P3: Evidence from a scalp EEG study', *Scientific Reports*, 5(1), p. 15129. doi: 10.1038/srep15129.

Li, S.-J. *et al.* (2002) 'Alzheimer Disease: evaluation of a functional MR imaging index as a marker.', *Radiology*, 225(1), pp. 253–259. doi: 10.1148/radiol.2251011301.

Lo, C. *et al.* (2010) 'Diffusion tensor tractography reveals abnormal topological organization in structural cortical networks in Alzheimer's disease.', *The Journal of neuroscience : the official journal of the Society for Neuroscience*, 30(50), pp. 16876–85. doi: 10.1523/JNEUROSCI.4136-10.2010.

Lopes da Silva, F. (1991) 'Neural mechanisms underlying brain waves: from neural membranes to networks', *Electroencephalography and Clinical Neurophysiology*, pp. 81–93. doi: 10.1016/0013-4694(91)90044-5.

Mehrkanoon, S. *et al.* (2014) 'Intrinsic Coupling Modes in Source-Reconstructed Electroencephalography', *Brain Connectivity*, 4(10), pp. 812–825. doi: 10.1089/brain.2014.0280.

Montez, T. *et al.* (2009) 'Altered temporal correlations in parietal alpha and prefrontal theta oscillations in early-stage Alzheimer disease', *Proceedings of the National Academy of Sciences*, 106(5), pp. 1614–1619. doi: 10.1073/pnas.0811699106.

Morabito, F. C. *et al.* (2015) 'A Longitudinal EEG Study of Alzheimer's Disease Progression Based on A Complex Network Approach', *International Journal of Neural Systems*, 25(2), p. 1550005. doi: 10.1142/S0129065715500057.

Mormino, E. C. *et al.* (2011) 'Relationships between beta-amyloid and functional connectivity in different components of the default mode network in aging', *Cerebral Cortex*, 21(10), pp. 2399–2407. doi: 10.1093/cercor/bhr025.

Mungas, D. (1991) 'In-office mental status testing: a practical guide.', *Geriatrics*, 46(7), pp. 54–8, 63, 66. Available at: http://www.ncbi.nlm.nih.gov/pubmed/2060803.

Nieuwenhuis-Mark, R. E. (2010) 'The death knoll for the MMSE: has it outlived its purpose?', *Journal of geriatric psychiatry and neurology*, 23(3), pp. 151–157. doi: 10.1177/0891988710375213.

O'Neill, G. C. *et al.* (2016) 'Measurement of Dynamic Task Related Functional Networks using MEG', *NeuroImage*. Elsevier, in press. doi: 10.1016/j.neuroimage.2016.08.061.

Onton, J. *et al.* (2006) 'Imaging human EEG dynamics using independent component analysis', *Neuroscience and Biobehavioral Reviews*, pp. 808–822. doi: 10.1016/j.neubiorev.2006.06.007.

de Pasquale, F. *et al.* (2010) 'Temporal dynamics of spontaneous MEG activity in brain networks.', *Proceedings of the National Academy of Sciences of the United States of America*, 107, pp. 6040–6045. doi: 10.1073/pnas.0913863107.

Pievani, M. *et al.* (2011) 'Functional network disruption in the degenerative dementias', *The Lancet Neurology*, pp. 829–843. doi: 10.1016/S1474-4422(11)70158-2.

Romei, V., Gross, J. and Thut, G. (2010) 'On the Role of Prestimulus Alpha Rhythms over Occipito-Parietal Areas in Visual Input Regulation: Correlation or Causation?', *Journal of Neuroscience*, 30(25), pp. 8692–8697. doi: 10.1523/JNEUROSCI.0160-10.2010.

Rubinov, M. and Sporns, O. (2011) 'Weight-conserving characterization of complex functional brain networks', *NeuroImage*, 56(4), pp. 2068–2079. doi: 10.1016/j.neuroimage.2011.03.069.

Sales-Pardo, M. *et al.* (2007) 'Correction for Sales-Pardo et al., Extracting the hierarchical organization of complex systems', *Proceedings of the National Academy of Sciences of the United States of America*, 104(47), p. 18874. Available at: http://www.pnas.org/cgi/content/abstract/104/39/15224.

Sarnthein, J. *et al.* (1998) 'Synchronization between prefrontal and posterior association cortex during human working memory.', *Proceedings of the National Academy of Sciences of the United States of America*, 95(12), pp. 7092–7096. doi: 10.1073/pnas.95.12.7092.

Sauseng, P. *et al.* (2010) 'Control mechanisms in working memory: A possible function of EEG theta oscillations', *Neuroscience and Biobehavioral Reviews*, pp. 1015–1022. doi: 10.1016/j.neubiorev.2009.12.006.

Schoffelen, J. M. and Gross, J. (2009) 'Source connectivity analysis with MEG and EEG', *Human Brain Mapping*, pp. 1857–1865. doi: 10.1002/hbm.20745.

Schroeder, C. E. and Lakatos, P. (2009) 'Low-frequency neuronal oscillations as instruments of sensory selection', *Trends in Neurosciences*, 32(1), pp. 9–18. doi: 10.1016/j.tins.2008.09.012.

Sheline, Y. I. *et al.* (2010) 'Amyloid Plaques Disrupt Resting State Default Mode Network Connectivity in Cognitively Normal Elderly', *Biological Psychiatry*, 67(6), pp. 584–587. doi: 10.1016/j.biopsych.2009.08.024.

Shirer, W. R. *et al.* (2012) 'Decoding subject-driven cognitive states with whole-brain connectivity patterns', *Cerebral Cortex*, 22(1), pp. 158–165. doi: 10.1093/cercor/bhr099.

Siegel, M., Donner, T. H. and Engel, A. K. (2012) 'Spectral fingerprints of large-scale neuronal interactions', *Nature Reviews Neuroscience*, 13(February), pp. 20–25. doi: 10.1038/nrn3137.

Sorg, C. *et al.* (2007) 'Selective changes of resting-state networks in individuals at risk for Alzheimer's disease.', *Proceedings of the National Academy of Sciences of the United States of America*, 104(47), pp. 18760–18765. doi: 10.1073/pnas.0708803104.

Sporns, O. (2010) *Networks of the brain*.

Stam, C. J. (2000) 'Brain dynamics in theta and alpha frequency bands and working memory performance in humans', *Neuroscience Letters*, 286(2), pp. 115–118. doi: 10.1016/S0304-3940(00)01109-5.

Stam, C. J. *et al.* (2009) 'Graph theoretical analysis of magnetoencephalographic functional connectivity in Alzheimer's disease', *Brain*, 132(1), pp. 213–224. doi: 10.1093/brain/awn262.

Stam, C. J. and Van Dijk, B. W. (2002) 'Synchronization likelihood: An unbiased measure of generalized synchronization in multivariate data sets', *Physica D: Nonlinear Phenomena*, 163(3–4), pp. 236–251. doi: 10.1016/S0167-2789(01)00386-4.

Stam, C. J. and van Straaten, E. C. W. (2012) 'The organization of physiological brain networks', *Clinical Neurophysiology*, pp. 1067–1087. doi: 10.1016/j.clinph.2012.01.011.

Von Stein, A. and Sarnthein, J. (2000) 'Different frequencies for different scales of cortical integration: From local gamma to long range alpha/theta synchronization', in *International Journal of Psychophysiology*, pp. 301–313. doi: 10.1016/S0167-8760(00)00172-0.

Supekar, K. *et al.* (2008) 'Network analysis of intrinsic functional brain connectivity in Alzheimer's disease', *PLoS Computational Biology*, 4(6). doi: 10.1371/journal.pcbi.1000100.

Tadel, F. *et al.* (2011) 'Brainstorm: A user-friendly application for MEG/EEG analysis', *Computational Intelligence and Neuroscience*, 2011. doi: 10.1155/2011/879716.

Tijms, B. M., Wink, A. M., *et al.* (2013) 'Alzheimer's disease: connecting findings from graph theoretical studies of brain networks', *Neurobiology of Aging*, pp. 2023–2036. doi: 10.1016/j.neurobiolaging.2013.02.020.

Tijms, B. M., Möller, C., *et al.* (2013) 'Single-Subject Grey Matter Graphs in Alzheimer's Disease', *PLoS ONE*, 8(3). doi: 10.1371/journal.pone.0058921.

Utianski, R. L. *et al.* (2016) 'Graph theory network function in parkinson's disease assessed with electroencephalography', *Clinical Neurophysiology*, 127(5), pp. 2228–2236. doi: 10.1016/j.clinph.2016.02.017.

Vecchio, F. *et al.* (2014) 'Human brain networks in cognitive decline: A graph theoretical analysis of cortical connectivity from EEG data', *Journal of Alzheimer's Disease*, 41(1), pp. 113–127. doi: 10.3233/JAD-132087.

Vecchio, F. *et al.* (2017) '"Small World" architecture in brain connectivity and hippocampal volume in Alzheimer's disease: a study via graph theory from EEG data', *Brain Imaging and Behavior*, 11(2), pp. 473–485. doi: 10.1007/s11682-016-9528-3.

Vemuri, P., Jones, D. T. and Jack Jr., C. R. (2012) 'Resting state functional MRI in Alzheimer's Disease', *Alzheimers Res Ther*, 4(1), p. 2. doi: 10.1186/alzrt100.

Wang, L. *et al.* (2006) 'Changes in hippocampal connectivity in the early stages of Alzheimer's disease: Evidence from resting state fMRI', *NeuroImage*, 31(2), pp. 496–504. doi: 10.1016/j.neuroimage.2005.12.033.

Watts, D. J. and Strogatz, S. H. (1998) 'Collective dynamics of "small-world" networks.', *Nature*, 393(6684), pp. 440–2. doi: 10.1038/30918.

World Health Organization (2012) 'Dementia: a public health priority', *Dementia*, p. 112. doi: 978 92 4 156445 8.

Zhao, X. *et al.* (2012) 'Disrupted small-world brain networks in moderate Alzheimer's disease: A resting-state fMRI study', *PLoS ONE*, 7(3). doi: 10.1371/journal.pone.0033540.

Zhou, J. *et al.* (2010) 'Divergent network connectivity changes in behavioural variant frontotemporal dementia and Alzheimer's disease', *Brain*, 133(5), pp. 1352–1367. doi: 10.1093/brain/awq075.